\documentclass[article,aps,prb,superscriptaddress, twocolumn]{revtex4-1}
\usepackage{amsmath}
\usepackage{natbib}
\usepackage{epsfig}
\usepackage{graphicx}
\usepackage{color}
\usepackage{multirow}
\usepackage{array}
\usepackage{tabularx}
\usepackage{inputenc}
\newcommand{\ymo}{YbMnO$_3$}

\newcommand{\rmo}{RMnO$_3$}
\newcommand{\y}{Yb$^{3+}$}
\newcommand{\m}{Mn$^{3+}$}
\newcommand{\re}{R$^{3+}$}

\begin{document}

\title{Single-crystal neutron diffraction study of hexagonal \ymo\ multiferroic \\ under magnetic field} 

\author{S. Chattopadhyay}
\email{s.chattopadhyay@hzdr.de}
\affiliation{Universit$\acute{e}$ Grenoble Alpes, INAC-MEM, F-38000 Grenoble, France}
\affiliation{Dresden High Magnetic Field Laboratory (HLD-EMFL), Helmholtz-Zentrum Dresden-Rossendorf, 01328 Dresden, Germany.}

\author{V. Simonet}
\affiliation{Institut N$\acute{e}$el, CNRS and Universit$\acute{e}$ Grenoble Alpes, BP166, 38042 Grenoble, France}

\author{V. Skumryev}
\affiliation{Departament de F\'isica, Universitat Aut\`onoma de Barcelona, 08193 Bellaterra, Barcelona, Spain}
\affiliation{Instituci\'o Catalana de Recerca i Estudis Avanc¸ats, E-08010 Barcelona, Spain}

\author{A. A. Mukhin}
\affiliation{Prokhorov General Physics Institute, Russian Academy of Sciences, 119991 Moscow, Russia}

\author{V. Yu. Ivanov}
\affiliation{Prokhorov General Physics Institute, Russian Academy of Sciences, 119991 Moscow, Russia}

\author{M. I. Aroyo}
\affiliation{Department of Condensed Matter Physics, University of the Basque Country UPV/EHU, Apartado 644, 48080 Bilbao, Spain}

\author{D. Z. Dimitrov} 
\affiliation{Institute of Solid State Physics, Bulgarian Academy of Sciences, 1184 Sofia, Bulgaria}
\affiliation{Institute of Optical Materials and Technologies, Bulgarian Academy of Sciences, 1113 Sofia,  Bulgaria}

\author{M. Gospodinov} 
\affiliation{Institute of Solid State Physics, Bulgarian Academy of Sciences, 1184 Sofia, Bulgaria}

\author{E. Ressouche}
\affiliation{Universit$\acute{e}$ Grenoble Alpes, INAC-MEM, F-38000 Grenoble, France}

\date{\today}

\begin{abstract}
We report single-crystal neutron diffraction study of the magnetic structure of the multiferroic compound \ymo, a member of the hexagonal manganite family, in zero-field and under a magnetic field applied along the $c$-axis. We propose a scenario for the zero-field magnetic ordering and for the field-induced magnetic reorientation of the Mn and of the two Yb on distinct crystallographic sites, compatible with the macroscopic measurements, as well as with previous powder neutron diffraction experiment and results from other techniques (optical second harmonic generation, M\"ossbauer spectroscopy). Our study should contribute in settling some debated issues about the magnetic properties of this material, as part of a broader investigation of the entire hexagonal \rmo\ (R = Dy, Ho, Er, Tm, Yb, Lu, Y) family.  
\end{abstract}

\maketitle{}


\section{Introduction}

\begin{figure}[!ht]
\includegraphics[angle=0,width=8.5 cm]{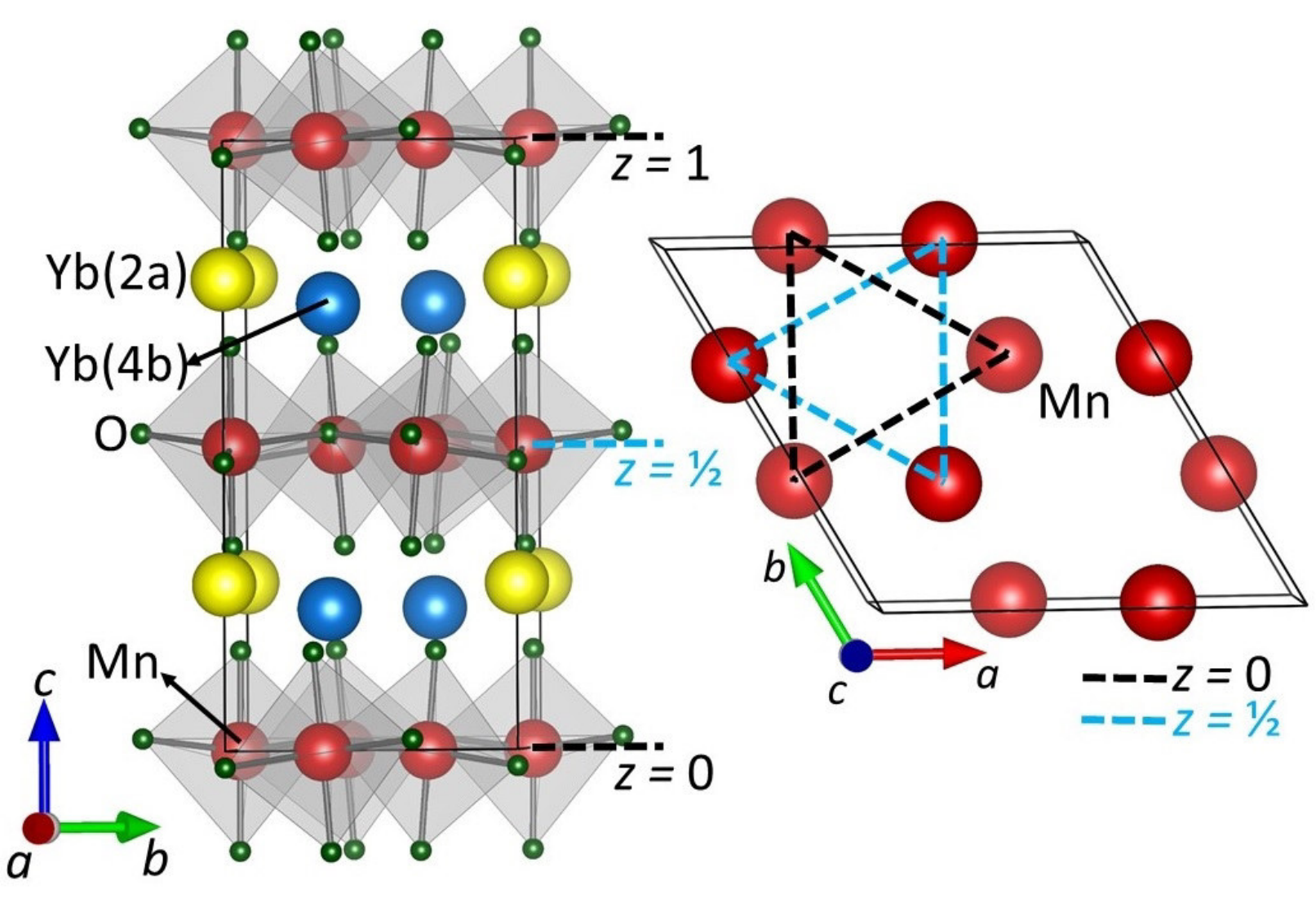}
\caption{Left panel: Perspective view of the \ymo\ crystal structure with tilted  MnO$_5$ bi-pyramids and Yb atoms on the different $4b$ and $2a$ Wyckoff sites respectively (the small green spheres are the oxygens). Right panel: Two layers of Mn triangles positioned at $z = 0$ and $z = 1/2$ along the crystallographic $c$ direction. }
\label{fig:str}
\end{figure}

The hexagonal h-\rmo\ multiferroic compounds (with R = Dy, Ho, Er, Tm, Yb, Lu, Y) have produced an abundant amount of literatures since their discovery in 1963 \cite{Bertaut63}. This interest is due to their exotic static and dynamical  behaviors ascribed to the combination of ferroelectricity and magnetic frustration. The latter arises from the intra-plane triangular arrangement of antiferromagnetically interacting Mn$^{3+}$ magnetic ions, weakly coupled along the $c$-axis, as shown in Fig. \ref{fig:str}. Unlike multiferroics where the ferroelectricity is induced by the magnetic order, as in the orthorhombic RMnO$_3$ compounds (with larger rare-earth ions), h-\rmo\ oxides become ferroelectric at much higher temperatures (around 1000 K) than their magnetic transition (N\'eel temperature $T_N$ below 100 K). In h-YMnO$_3$, ferroelectricity was found to be connected to the buckling of the layered MnO$_5$ polyhedra, displacements of the Y ions, and the trimerization of the Mn lattice associated to strong magnetoelastic effects \cite{Aken04,Fennie05, Lee08, Lilienblum15,Sim16,Ban18,Sim18,Skg18}. Although the exact mechanism at the origin of the ferroelectricity has been debated, it is assumed to be identical in all members of the family, which are described in the hexagonal space group $P6_3cm$ (number 185) at low temperature. 

\begin{figure}[!ht]
\includegraphics[angle=0,width=8.5cm]{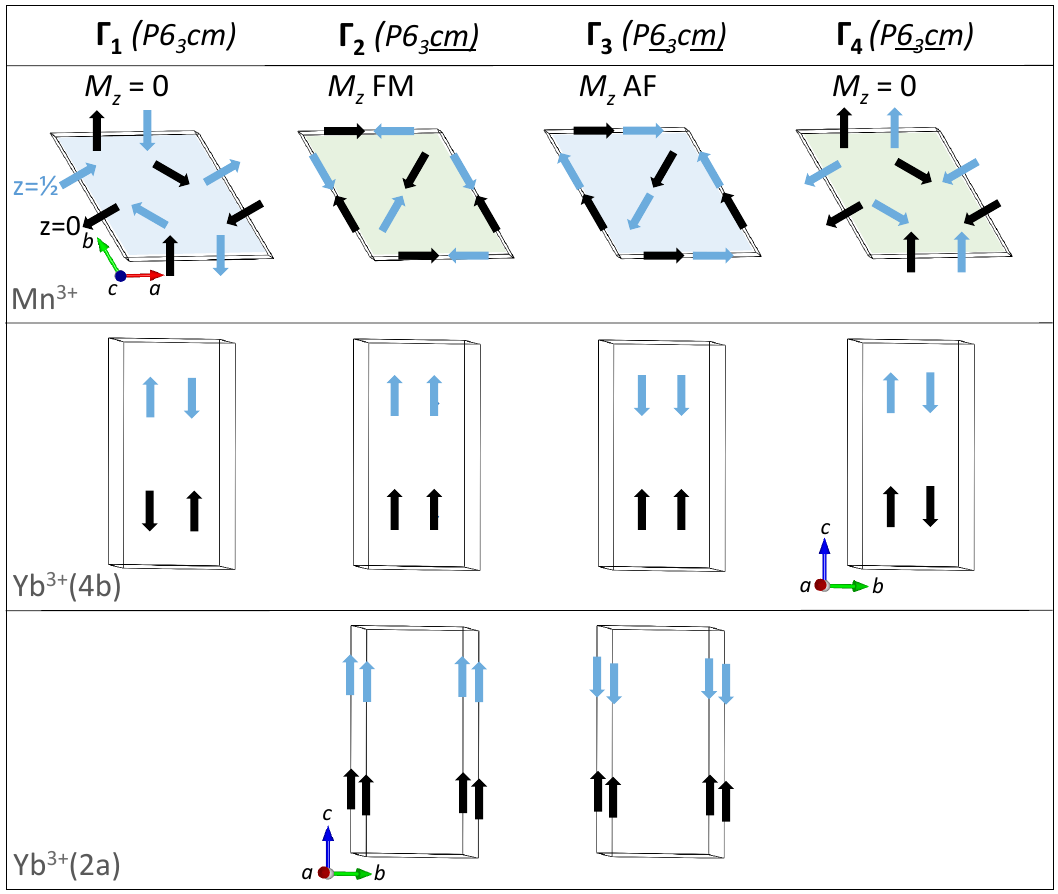}
\caption{Magnetic structures associated to the one-dimensional irreducible representations of $P6_3cm$ ($k$=0) for \m, \y($4b$), and \y($2a$) shown at the top, middle, and bottom panels respectively. The corresponding isotropy magnetic groups are also indicated. The Mn homometric pairs are indicated by similar color fillings. The magnetic moments in the $z=0$ and $z=1/2$ planes are shown in black and blue respectively. For the \m, the $z$ component of the magnetic moment with its ferromagnetic (FM) or antiferromagnetic (AF) coupling is also indicated.}
\label{fig:ir}
\end{figure}

The magnetism is a complex issue in itself in this class of materials. \m\ ions occupy the Wyckoff site $6c$, forming triangular layers of \m\ in the ($a$, $b$) planes. Between them, the \re\ occupies two different crystallographic Wyckoff sites $4b$ and $2a$. While decreasing the temperature, first the \m\ order magnetically at T$_N$ due to superexchange antiferromagnetic interactions in a 120$^{\circ}$ magnetic structure, characterized by a $k$=(0, 0, 0) propagation vector \cite{Bertaut63}. Then the \re($4b$) are polarized in the molecular field of the Mn, while the \re($2a$) are believed to order at much lower temperature from their mutual interactions \cite{Sugie02}. Additional spin-reorientations occur at intermediate temperatures for some members of the family ({\it e.g.} Ho, Sc). Finally, metamagnetic processes are frequently observed under magnetic field. Magnetoelectric coupling is also evidenced from strong dielectric anomalies visible at each magnetic transitions \cite{Hur09}. The major tools that have been used to determine the various magnetic configurations of h-\rmo\ are neutron diffraction and optical second harmonic generation (SHG). They often but not always agree. For the unpolarized neutron scattering method, difficulties arise from the  indetermination among different possible magnetic structures (homometric pairs) \cite{Brown06}, whereas the SHG technique has difficulty to distinguish different magnetic sublattices. In addition, neutron scattering has often been performed only on polycrystalline samples which is usually not sufficient to determine field-induced magnetic structures. 
    
In this article, we concentrate on \ymo, which was reported to undergo a magnetic transition below $T_N$ $\sim$ 80 K with a propagation vector $k = (0, 0, 0)$ \cite{Sugie02,Fontcuberta08}. Powder neutron diffraction \cite{Fabreges08} and SHG \cite{Fiebig00,Fiebig03} agree on the high temperature magnetic ordering. However the interpretation of the low temperature magnetic configuration, in particular when the Yb($2a$) are expected to play a role, is not clear, as is the exact mutual orientations of the three magnetic ions under magnetic field. Moreover, hysteretic effects, the coexistence of competing phases even in zero field or a spin reorientation at intermediate temperature have also been reported \cite{Fiebig03,Adem09,Qiang16} justifying the necessity of single-crystal neutron diffraction experiment on this composition.


\section{Synthesis and Experimental Details}

The plate-like single crystals of \ymo, with the hexagonal $c$-axis perpendicular to the surface and a thickness of about 0.5 mm, were grown using flux method as described by Yen {\it et al} \cite{Yen07}. The magnetization ($M$) measurements were performed in a commercial SQUID magnetometer and  in a physical property measurement system from Quantum Design. Electric polarization ($P$) was measured along the crystallographic $c$-axis using a Keithley 6517A electrometer. Electrical contacts were attached by Ag paint on the two parallel surfaces of the $c$-cut sample. The measurements were performed both for zero and non-zero ($\pm$1~kV/mm) applied electric field regimes. These have not revealed a significant difference, implying the absence of ohmic currents. Note that the latter were observed in YMnO$_3$ at temperatures above 230 K and were accompanied by a negative magnetoresistive effect \cite{Huang97}. No preliminary poling was carried out to polarize the sample possessing a ferroelectric domain structure. Neutron diffraction measurements in zero magnetic field were performed on the CEA-CRG D15 and D23 single crystal diffractometers at the ILL (wavelength $\lambda$=1.173 and 1.27 \AA\ respectively) in four-circle mode using a standard orange cryostat. The CEA-CRG D23 diffractometer at the ILL was also used for the measurements under magnetic field using a lifting arm detector and a vertical cryomagnet. The zero-field measurements on D15 and D23 were checked to be consistent and results coming out of D23 only are presented in the following. 

\begin{figure}[!ht]
	\includegraphics[angle=0,width=8.5cm]{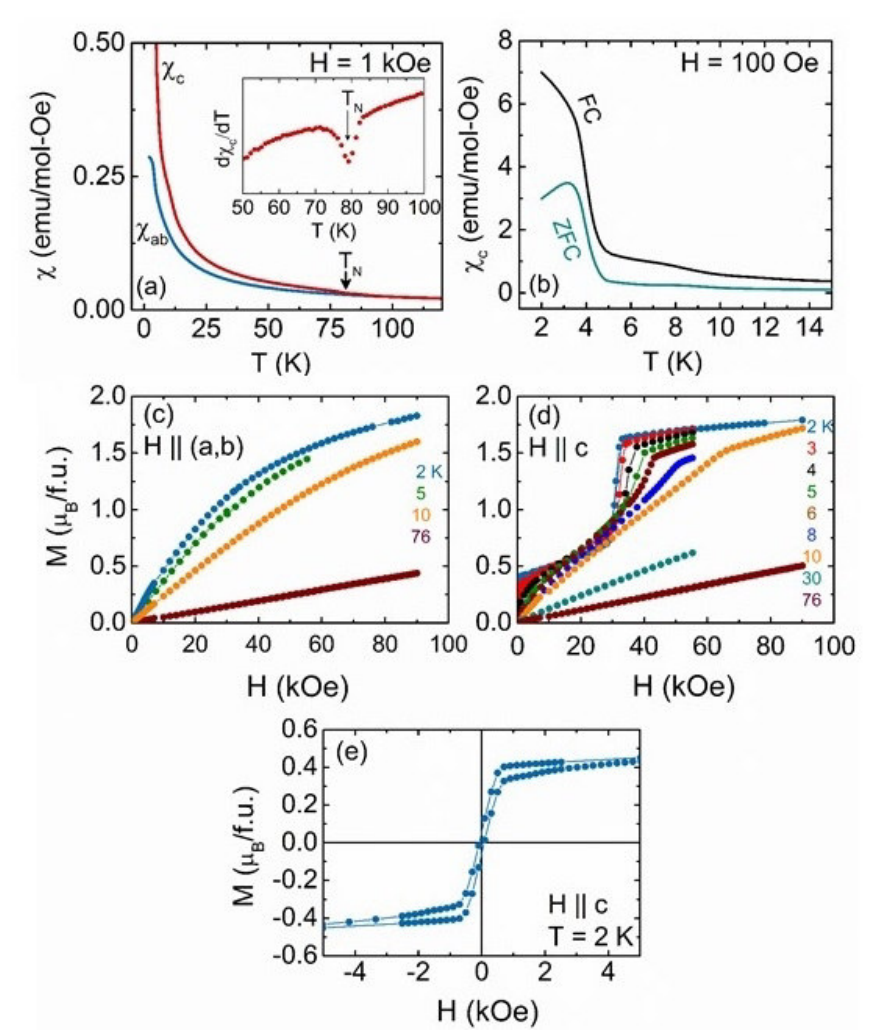}
	\caption{(a) DC magnetic susceptibility ($\chi$) along $c$-axis and in the $(a,b)$ plane. Measurements were performed in a magnetic field of 1 kOe on cooling. Inset: $\frac{d\chi_c}{dT}$ vs. $T$ to show the transition at $T_N$ = 80 K. (b) $\chi$($T$) measured with a field of 100 Oe along the $c$-axis after field cooling and zero field cooling procedures, showing the low temperature anomaly associated to the \y($2a$) ordering. Magnetization ($M$) vs. applied magnetic field ($H$) for $H$ in the ($a$, $b$) plane (c) and along $c$ (d). The low-field $M$($H$) curve at 2 K for $H \parallel c$ is shown in panel (e). The data are not corrected for demagnetizing field effect.}  
	\label{fig:mag}
	
\end{figure}

\begin{figure}[t]
	\includegraphics[angle=0,width=7.5cm]{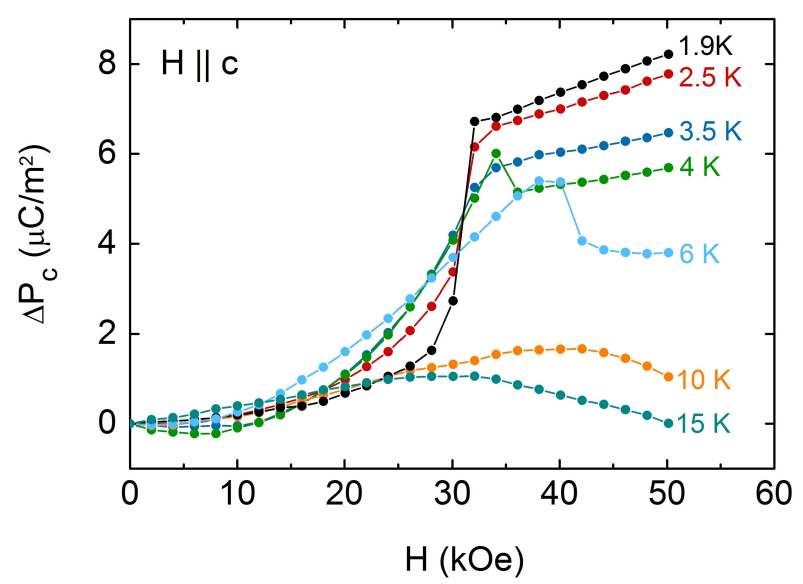}
	\caption{Change of electric polarization ($\Delta P_c$) along the $c$-axis as a function of magnetic field ($H$) applied along the same direction (measured in an applied electric field of 1~kV/mm).}
	\label{fig:pol}
\end{figure}



\section{Irreducible Representation Analysis}   

Let us recall that Group theory and representation analysis \cite{Bertaut68} reveal that the representation of the magnetic structure for \m and \y($4b$) involve six possible irreducible representations (IR) denoted as $\Gamma_i$ ($i$ = 1,2,...,6), which are compatible with the $P6_3cm$ space group and the propagation vector $k$ = (0, 0, 0) \cite{Munoz00}. Four of these, $\Gamma_1$ to $\Gamma_4$, are unidimensional and depicted in Fig. \ref{fig:ir}, the last two ones being two-dimensional. For the \y($2a$) site, only the $\Gamma_2$, $\Gamma_3$, $\Gamma_5$, and $\Gamma_6$ irreducible representations are allowed. 

Concerning the \m\ site, the magnetic moments are constrained in the $(a,b)$ plane for $\Gamma_1$ and $\Gamma_4$, while an out-of-plane ferromagnetic component is allowed for $\Gamma_2$, and an out-of-plane antiferromagnetic component is allowed for $\Gamma_3$. Among these configurations, those of symmetry $\Gamma_1$ and $\Gamma_3$ correspond to homometric pairs, so as $\Gamma_2$ and $\Gamma_4$. This means that these pairs are hardly distinguishable (almost identical intensities of the magnetic Bragg reflections) by neutron diffraction when the \m\ coordinate $x_{Mn}$ is close to 1/3. For the two-dimensional representations $\Gamma_5$ and $\Gamma_6$, the Fourier components of the magnetic moments are written as a linear combination of 6 basis functions. The corresponding magnetic structures are reported in ref. \onlinecite{Munoz00}. For $\Gamma_5$, there are four magnetic modes with the magnetic moments in the ($a$, $b$) plane, displaying either a ferromagnetic or a 120$^{\circ}$ arrangement. The coupling is ferromagnetic between the planes. The two other modes are discarded since they concern non equal moments along the $c$-axis. Similar solutions are found for $\Gamma_6$ but with an antiferromagnetic coupling of the in-plane magnetic structures along the $c$-axis. Concerning the \y\ $4b$ and $2a$ sites, the one dimensional IR correspond to magnetic moments along the $c$-axis.
Importantly, only $\Gamma_2$ configuration allows a ferromagnetic component along the $c$-axis for the three sites (cf. Fig. \ref{fig:ir}).

\section{Results}


\subsection{Magnetization and electric polarization}

Fig. \ref{fig:mag}(a) depicts the temperature ($T$) variation of the dc magnetic susceptibility $\chi$ in magnetic field of 1 kOe along the $c$ direction ($\chi_c$) and within the $(a,b)$ plane ($\chi_{ab}$), recorded on cooling. Anisotropic response is seen, apparently below the N\'eel temperature, with $\chi_c$/$\chi_{ab}$ ratio value reaching  $\sim$15 at 2 K. The data well replicate the results reported in previous literature \cite{Sugie02,Yen07,Fontcuberta08,Fabreges08,Skumryev09}. The anomaly near 80 K ($T_N$) in the $\chi_c$ data (Inset of Fig. \ref{fig:mag}(a)) corresponds to the ordering of \m\ magnetic moments as reported earlier. The sharp increase of $\chi_c$ below about 5 K (Fig. \ref{fig:mag}(b)) signals the onset of long range ordering of the \y($2a$) moments. The differences in the zero field and field cooled thermomagnetic curves below about 4 K reflect the existence of net magnetic moment and coercivity. 

The field dependence of the magnetization for various temperatures are shown in Fig. \ref{fig:mag}(c) and Fig. \ref{fig:mag}(d) for a field applied perpendicular and parallel to the $c$-axis respectively, and is again consistent with previous reports \cite{Sugie02,Yen07, Abramov11,Midya11,Lorenz13}. No anomaly, nor hysteretic behavior, are measured for the magnetic field applied in the ($a$, $b$) plane. This is at variance with the behavior observed for the other orientation of the field where two step-like features are visible in the magnetization curve at low temperature. The  low field magnetization step is observed only for temperatures below about 5 K. As seen on the 2 K curve (Fig. \ref{fig:mag}(e)), the magnetization reaches $\sim$0.4 $\mu_B$/f.u. almost instantly after increasing the magnetic field (as soon as the applied field reaches $H_{c1}$$\sim$ 300 Oe, which means that the internal field is even smaller). The magnetization step value as well as the magnetization hysteresis seen on Fig. \ref{fig:mag}(e) point to ferrimagnetic alignment between \y($4b$) and \y($2a$) moments. 
The second step like anomaly appears near $H_{c2}$ = 30 kOe for the 2 K curve. At this field, a jump of magnetization takes place and it increases with about 1.2 $\mu_B$/f.u. above its low field step value of $\sim$0.4 $\mu_B$/f.u. and becomes closer to the ($a$, $b$) plane magnetization.  This step-like feature actually reveals a {\it field induced} spin reorientation that has been addressed through our neutron diffraction measurements presented below. It smears out with increasing temperature along with the shift of $H_{c2}$ towards higher field.

The field dependence of the electric polarization change $\Delta P_c$($H$) along the $c$-axis is presented in Fig. \ref{fig:pol} for several temperatures. The magnetic field was applied along the same direction. A sharp jump of the polarization is observed at $\sim$30 kOe at the lowest temperatures. It is clearly correlated to the field-induced magnetic phase transition since its threshold field coincides with the second step-like anomaly in the magnetization curves at $H_{c2}$. The polarization anomaly smears out with increasing temperature. Its field-dependence changes character between $T_c\sim$~4 K and 6 K revealing a broad maximum in $\Delta P_c$($H$) at higher temperatures. These modifications are probably associated to a change in the nature of the field-induced transition above and below the \y($2a$) ordering at $T_c$.

\begin{figure}[!ht]
\includegraphics[angle=0,width=7.5cm]{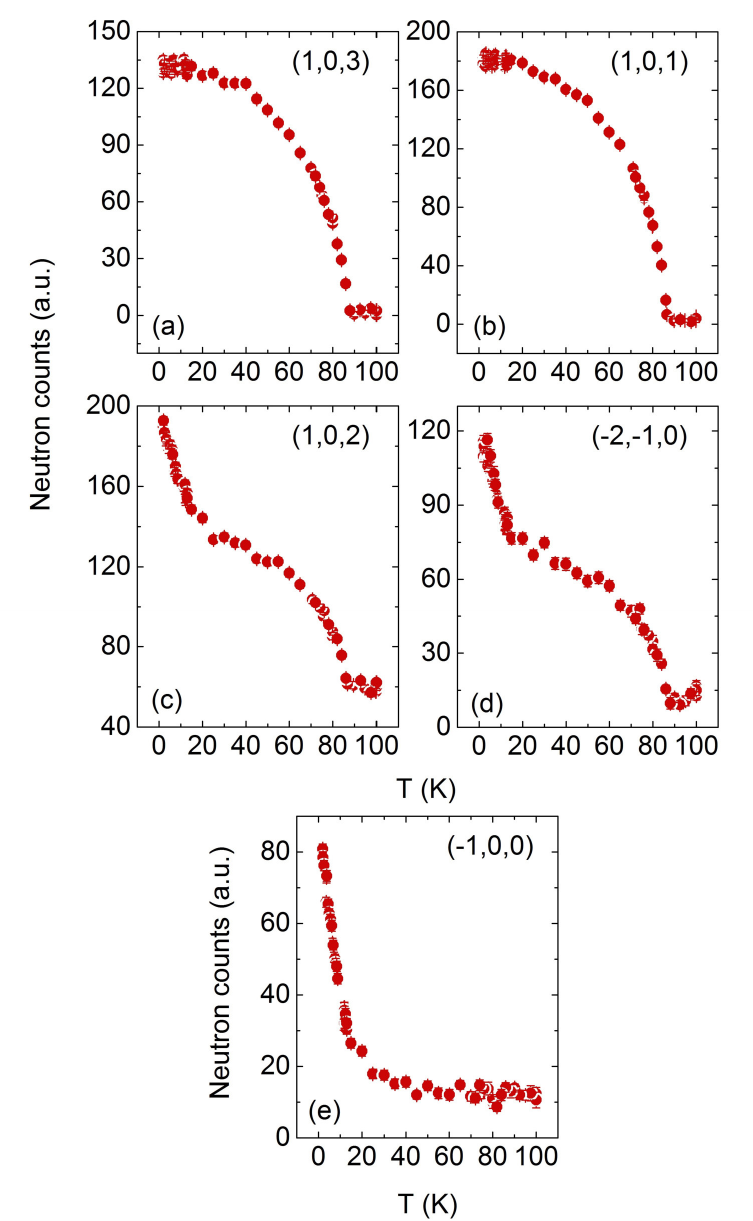}
\caption{Neutron counts at the peak maximum versus temperature for three types of Bragg reflections. They are site-specific for the magnetic arrangements corresponding to $\Gamma_2$ and $\Gamma_4$: (a-b) (1, 0, 3) and (1, 0, 1) contributed solely by ordered \m\, (c-d) (1, 0, 2) and (-2, -1, 0) contributed by all \m, \y($4b$) and \y($2a$), and (e) (-1, 0, 0) contributed solely by \y($4b$) and \y($2a$). Note that $\Gamma_4$ configuration is not allowed by symmetry for the \y($2a$) magnetic order.}
\label{fig:it}
\end{figure}

\begin{figure}[!ht]
	\includegraphics[angle=0,width=8.5cm]{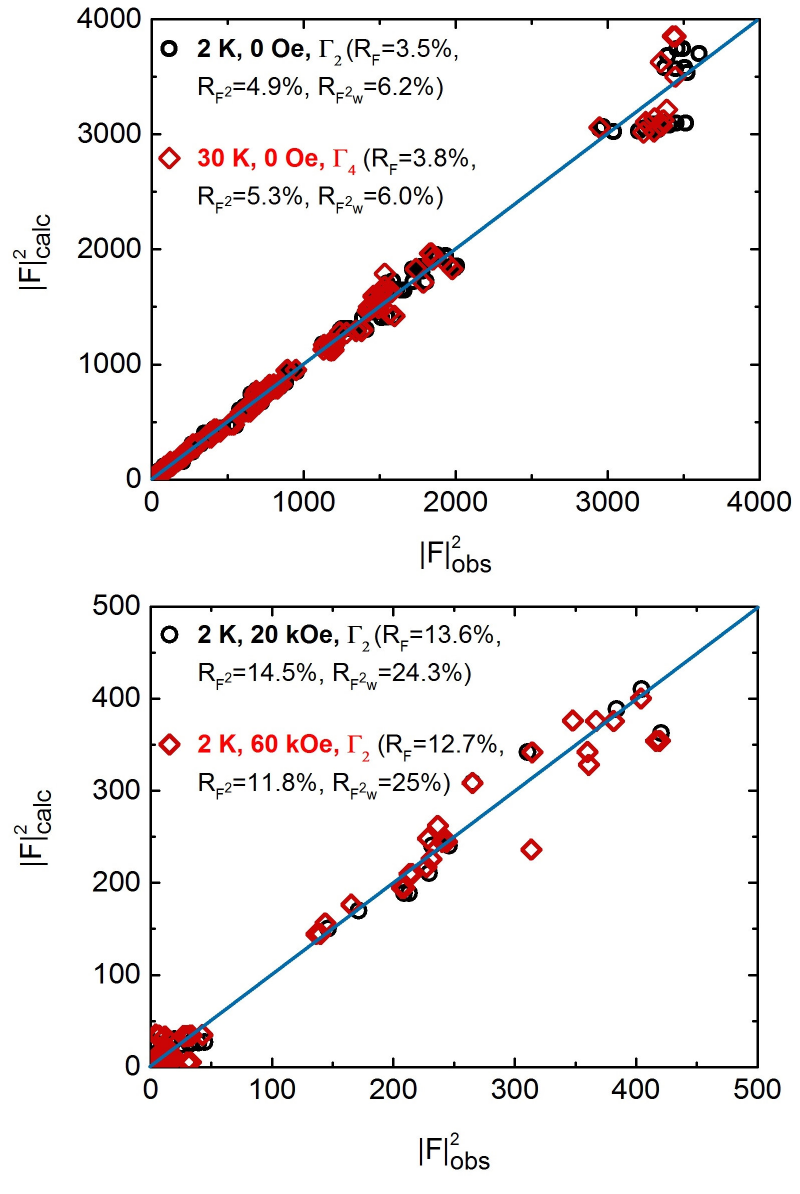}
	\caption{\small{Observed versus calculated intensity (square of the structure factor) of the magnetic Bragg reflections in arbitrary units. Top panel: Data recorded in zero field at $T$ = 30 K (red diamond) and 2 K (back circle) using the 4-circles mode. Bottom panel: Data recorded at $T$ = 2 K with $H$ = 20 kOe (black circle) and 60 kOe (red diamond) in the cryomagnet using the lifting-arm detector. The agreement factors and magnetic configurations for the three sites are indicated.}}
	\label{fig:refin}
\end{figure}

\begin{table}[t]
	
	\caption{\small{Results of the refinement of the single crystal neutron diffraction Bragg peak intensities recorded above (100 K) and below (30 K, 13 K, and 2 K) the transition temperature ($T_N$ = 80 K) under zero magnetic field condition. $R_F$, $R_{F^2}$, and $R_{F^2w}$ are the agreement factors of the fits \cite{Rodriguez93}, $M$ is the magnetic moment, and $x_{\rm Mn}$ is the $x$ coordinate of the \m.}}
	
	\label{table4}
	
	\begin{tabular} {ccc}
		
		\hline\\
		
		$T$ = 100 K, $x_{\rm Mn}$ = 0.3345(20)\\
		($R_F$ = 4.7\%, $R_{F^2}$ = 5.5\%, $R_{F^2w}$ = 6.3\%)\\
		\hline\\
		$T$ = 30 K, $x_{\rm Mn}$ = 0.3343(14)\\
		($R_F$ = 3.8\%, $R_{F^2}$ = 5.3\%, $R_{F^2w}$ = 6.0\%)\\
		
		Ion/IR & $M$ ($\mu_B$) \\
		\m/$\Gamma_4$ &3.32(3)  \\
		\y($4b$)/$\Gamma_4$ &0.81(5) \\
		
		\hline\\
		$T$ = 13 K, $x_{\rm Mn}$ = 0.3350(15) \\
		($R_F$ = 3.9\%, $R_{F^2}$ = 5.3\%, $R_{F^2w}$ = 6.4\%)\\
		
		Ion/IR & $M$ ($\mu_B$) \\
		\m/$\Gamma_4$ &3.41(4) \\
		\y($4b$)/$\Gamma_4$ &1.15(5)  \\
		
		\hline\\
		$T$ = 2 K, $x_{\rm Mn}$ = 0.3349(15) \\
		($R_F$ = 3.5\%, $R_{F^2}$ = 4.9\%, $R_{F^2w}$ = 6.2\%)\\
		
		Ion/IR & $M$ ($\mu_B$) \\
		\m/$\Gamma_2$ &3.41(3) \\
		\y($4b$)/$\Gamma_2$ &1.76(7)  \\
		\y($2a$)/$\Gamma_2$ &-1.47(8)  \\
		
		\hline\\
		
	\end{tabular}
	
\end{table}

\begin{figure*}[!ht]
\includegraphics[angle=0,width=14 cm]{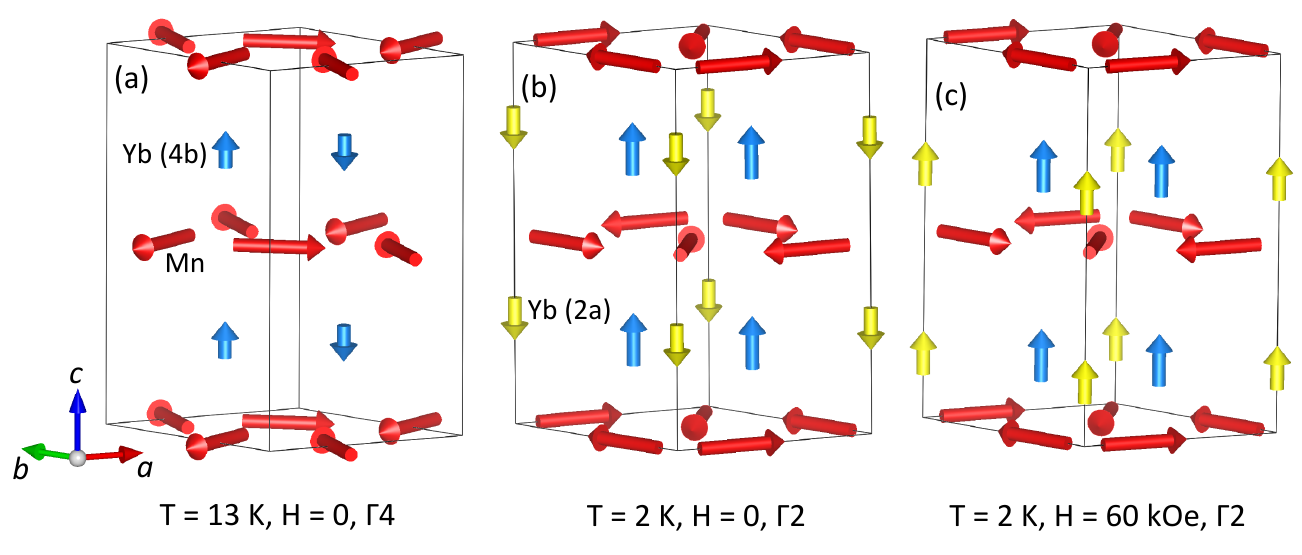}
\caption{Magnetic structures (and corresponding IR) refined in zero magnetic field at (a) 13 K and (b) 2 K. (c) Magnetic structure inferred from the neutron and magnetization data in 60 kOe field  at 2 K. The \m\ moments are represented in red, the \y($4b$) in blue, and the \y($2a$) in yellow.}
\label{fig:magstruc}
\end{figure*}
  

\subsection{Single-crystal neutron diffraction}  


\subsubsection{In zero magnetic field}

About 700 Bragg reflections were recorded at 100 K, 30 K, 13 K, and 2 K (the magnetic Bragg peaks associated to the $k$=(0, 0, 0) propagation vector rise on top of the nuclear ones). All the refinements were performed using the FullProf Suite software package \cite{Rodriguez93} and we used the formalism of Becker-Coppens to refine the anisotropic extinction \cite{Becker74}. The refined atomic positions of the 100 K data (above $T_N$) were found very similar to the ones reported from X-ray diffraction at room temperature \cite{Isobe91,Aken01}. In particular, the $z$ coordinate of \m\ is almost 0 and thus kept to zero for the lower temperature refinements. The $x$ coordinate of the \m\ ($x_{\rm Mn}$) is also very close to 1/3  (see Table I) but it was allowed to vary in the lower temperature refinements as its variation is believed to be intricately linked to the selection of the magnetic order \cite{Fabreges09}. 

To get a deeper insight on the contribution of the different sublattices to the magnetic order in zero field, we chose to study the thermal evolution of some characteristic site-specific reflections (see Fig. \ref{fig:it}). We followed the $(1, 0, 1)$ and $(1, 0, 3)$ reflections, which are forbidden in the $P6_3cm$ space group. As can be seen, they start to become non-zero only below $T_N$ reflecting their magnetic character and have a temperature dependence compatible with that of a magnetic order parameter. Under $\Gamma_2$ and $\Gamma_4$ symmetry, those two reflections are entirely dependent on the ordering of \m\ magnetic moments. Therefore, the rise of these reflections below $T_N$ is consistent with the coincidence of the N\'eel temperature with the magnetic ordering of the \m\ in one of these two IR. In contrary, the $(-1,0,0)$ reflection shown in Fig. \ref{fig:it}(e) only depends, for its magnetic component, on the \y\ magnetic moments for $\Gamma_2$ ($2a$ and $4b$ sites) and $\Gamma_4$ (only $4b$ site). The intensity of this reflection increases very slowly when decreasing the temperature down to $\sim$20-30 K, below which, it shows a sharp increment. Such a behavior is consistent with the magnetic ordering of the \y($4b$) in the \m\ molecular field \cite{Fabreges08}. Note that, as the transition at $T_N$ is second order, the two \m\ and \y($4b$)\ sublattices must order with the same IR ($\Gamma_2$ or $\Gamma_4$) since they are coupled. No anomaly can be seen in $(-1,0,0)$ at $T_c$. Two other characteristic reflections viz. $(1,0,2)$ and $(-2,-1,0)$ combine the features of the previous two kinds of reflections, thus involving magnetic contributions from both \m\ and \y.

We checked the information deduced from the temperature dependence of selected reflections through the refinement of all the Bragg reflections below $T_N$. To obtain the best refinements, all the possible one-dimensional IR as predicted for \m\ and \y\ were tested. In addition to the magnitude of the magnetic moments, the $x$ coordinate of the \m\ ions were also refined. At 30 and 13 K, the best refinements with similar agreement factors were obtained for the representations $\Gamma_4$ or $\Gamma_2$ for \m/\y($4b$) (homometric \m\ IR), with an enhancement of the \y($4b$) moment from 30 to 13 K and no magnetic moment on the \y($2a$) sites. However, in the $\Gamma_2$ magnetic configuration, the \y($4b$) are ferromagnetically coupled. From the neutron refinement, they should give rise to a ferromagnetic contribution amounting to $\sim$0.7 $\mu_B$ at 30 K and $\sim$0.9 $\mu_B$ at 13 K, which is not observed in the magnetization measurements. A small bifurcation of the field-cooled and zero-field-cooled susceptibilities has been reported \cite{Fontcuberta08} below $T_N$ but it cannot be accounted by such a large magnetization and is rather due to field-polarized defective magnetic moments for instance in the domain walls \cite{Yen07}. Our finding thus points out to the $\Gamma_4$ magnetic configuration for \m\ and \y($4b$) (see Figs. \ref{fig:refin} and \ref{fig:magstruc}), which is consistent with previous powder neutron diffraction \cite{Fabreges08} and SHG results \cite{Fiebig00}. The good quality of the fits obtained using $\Gamma_4$/$\Gamma_2$ configurations allowed us to limit our analysis to solutions corresponding to one-dimensional IR over the more complex two-dimensional ones.

M\"ossbauer and far-infrared spectroscopies have proven the ordering of the \y($2a$) below $\sim$5 K \cite{Fabreges08,Salama09,Standard12}. Moreover, the onset of a ferromagnetic component along the $c$-axis is associated to this ordering, which suggests that the \y($2a$) moments order in the $\Gamma_2$ IR. The last uncertainty concerns the ordering of the \m\ and \y($4b$) below $T_c$. Do they remain in the  $\Gamma_4$ IR or do they reorient due to a coupling with the \y($2a$) in the $\Gamma_2$ IR? We checked both possibilities: Although the agreement factor is only slightly better for the $\Gamma_2$ solution for all sites versus the solution with $\Gamma_4$ for \m\ and \y($4b$) and $\Gamma_2$ for \y($2a$), the former is more consistent with the magnetization data yielding an almost 0.5 $\mu_B$ per formula unit as shown in Fig. \ref{fig:mag}(d). Indeed, the total ferromagnetic component deduced from the neutron refinement for the $\Gamma_2$ solution is found equal to 0.68(8) $\mu_B$ per formula unit, instead of 0.06(3) $\mu_B$ obtained for the $\Gamma_4$/$\Gamma_2$ solution. We have also tested the possibility of an in-plane configuration of \y($2a$) magnetization that had been suggested \cite{Fabreges08} but found a worse agreement with our data. The results of the fits are shown in Fig. \ref{fig:refin} with the associated magnetic configuration in Fig. \ref{fig:magstruc} and the details of the important refined parameters are given in Table I.


\begin{figure}[!ht]
	\includegraphics[angle=0,width=8.5cm]{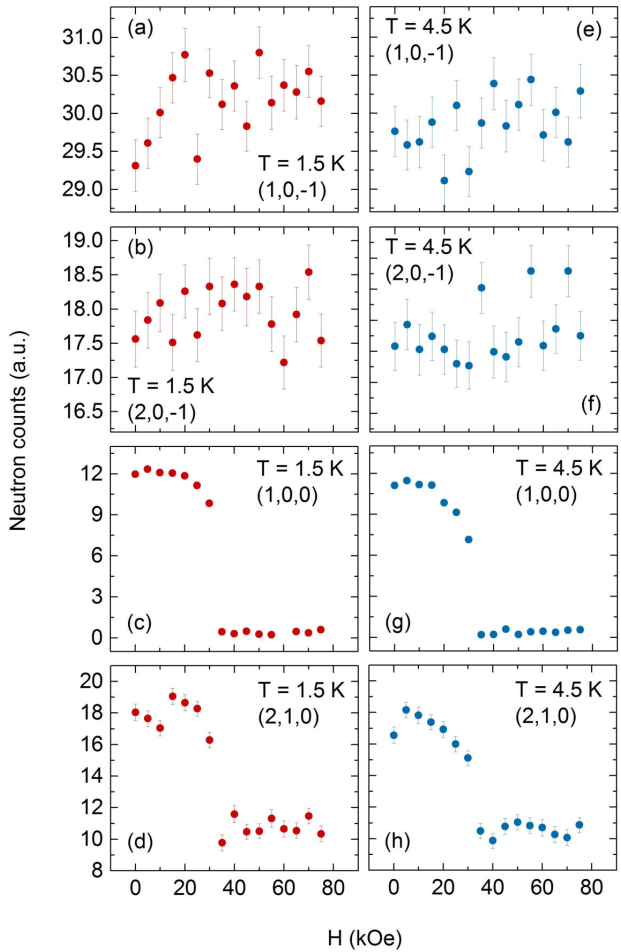}
	\caption{Neutron counts at the peak maximum versus magnetic field at 1.5 K (left) and 4.5 K (right) for different types of Bragg reflections that are site-specific in the 2$^{\rm nd}$ and 4$^{\rm th}$ IRs for their magnetic contribution. }
	\label{fig:ih}
\end{figure}

\subsubsection{Under magnetic field}

Fig. \ref{fig:ih} depicts the magnetic field dependence of characteristic reflections measured at 1.5 K and 4.5 K, i.e. below and close to the transition $T_c$, with a field applied along the $c$-axis. The results are identical: The $(1, 0, -1)$ [also $(-2, 0, 1)$] reflection, which depends solely on \m\ under the $\Gamma_2$ IR, remains practically insensitive to the field sweep. On the other hand, the $(1, 0, 0)$ reflection, which is associated with \y($4b$) and \y($2a$) magnetic moments in the $\Gamma_2$ IR, strongly decreases at 30 kOe and almost vanishes above this field. This field corresponds to $H_{c2}$ marking the high field step-like anomaly observed in the macroscopic measurements for a field applied along the $c$-axis. The last reflection (2, 1, 0) combines the two contributions from the \m\ and from the \y. It is therefore evident that only the \y\ moments are affected by the application of an external magnetic field. 

Although nearly 140 reflections were recorded at 2 K for $H$ = 0.25 kOe ($H < H_{c2}$), and 60 kOe ($H > H_{c2}$), the small coverage of the reciprocal space due to the limited aperture of the cryomagnet did not allow us to perform a reliable refinement of the data. Instead, we tested the most reasonable assumption agreeing with the magnetization data: a reorientation of the \y($2a$) magnetic moments above $H_{c2}$ in a way that they become parallel to the \y($4b$) ones while keeping the \m\ magnetic moments unaltered in the ($a$, $b$) plane. This reorientation would lead, keeping the \y($4b$) and \y($2a$) magnetic moments to their absolute values refined in zero field, to a total magnetization per f.u. equal to 1.65 $\mu_B$ in agreement with the magnetization data of Fig. \ref{fig:mag}. As starting point for the analysis of the data under field, we tested the magnetic configuration found previously in zero-field using the 4-circles mode. This configuration yields comparable agreement factors using the data recorded with the cryomagnet and lifting-arm detector in zero field and under a magnetic field of 20 kOe ($R_F/R_{F^2}/R_{F^2w}$ = 13.7/15.4/18.7\% and 13.6/14.5/24.3\% respectively). At 60 kOe, i.e. above $H_{c2}$, the same configuration does not hold good any longer as evidenced by worse agreement factors ($R_F/R_{F^2}/R_{F^2w}$= 24.2/17.9/43.8\%). Those improve significantly ($R_F/R_{F^2}/R_{F^2w}$ = 12.7/11.8/25\%) by reversing the \y(2$a$) magnetic moments and making them parallel to the \y($4b$) ones (see bottom panel of Fig. \ref{fig:refin}), hence validating the model. It is to be noted that the strong decrease of the $(1, 0, 0)$ Bragg reflection above $H_{c2}$, as depicted in Fig. \ref{fig:ih}, is obtained within this model (where all the \y\ magnetic moments are aligned along the field) due to the destructive interference between the two \y\ sublattices. 

At 10 K ($T_c<T<T_N$) the same set of reflections was recorded at field values of 4 T and at 8 T, i.e. below and above the critical field corresponding to $H_{c2}$ for this temperature (see Fig. \ref{fig:mag}(c)). Again, a full refinement is impossible but the data unambiguously show that the \y($4b$) and \y($2a$) moments order in the $\Gamma_2$ IR and are aligned parallel to each other and to the field for both values of the field. It is impossible to determine whether the \m\ magnetic moments are in $\Gamma_2$ or $\Gamma_4$. At 8 T, the data are compatible with the same configuration than the one inferred from that data at 2 K and 6 T which corresponds to the full polarization of the two \y\ sites. Note that the low field phase with both \y\ moments antiparallel to each other seems not to exist at 10 K.


\subsection{Summary and discussion}

In summary, we have achieved a good description of the magnetic properties of the h-YbMnO$_3$ compound through our single-crystal neutron diffraction study. We have found that the \m\ magnetic moments order below $T_N$ = 80 K, polarizing the \y($4b$) moments whose ordered component strongly increases below 20-30 K. Both \m\ and \y($4b$) moments are described with respect to $\Gamma_4$ in this temperature region. Below 5 K, the \y($2a$) moments order in $\Gamma_2$ dragging the \m\ and \y($4b$) into a new kind of magnetic configuration also corresponding to $\Gamma_2$. It consists of a  ferromagnetic arrangement of the \y($2a$) and \y($4b$) moments along $c$ in a way that the \y($2a$) and \y($4b$) magnetic sublattices are antiparallel to each other. There happens also to be an important spin-reorientation of the in-plane \m\ moments by 90$^{\circ}$, as well as an additional change of its interlayer coupling from AFM to FM. Under a magnetic field applied along the $c$-axis, our magnetization and neutron diffraction data are compatible with a spin-flip of the \y($2a$) moments that become aligned with the field and with the \y($4b$) magnetic moments. This correlates well with the observed step-like change in the electric polarization accompanying the field-induced magnetic transition. These polarization changes could originate from small alterations in the atomic/electronic positions due to spin-lattice coupling together with a change in Yb-Mn and Yb-Yb magnetic interactions. 

The $H-T$ phase diagram described above is consistent with the SHG results and very similar to the one reported for the h-ErMnO$_3$ compound \cite{Meier12}. It has been rationalized phenomenologically through the Laudau theory of phase transition \cite{Munawar06}. An important microscopic parameter seems to be the Mn-rare earth coupling. In the present study, it is evidenced through the polarization of the \y($4b$) moments by the \m\ ones and through the reorientation of the \m\ moments triggered by the \y($2a$) magnetic ordering. This coupling has also signatures in the dynamical properties of YbMnO$_3$ \cite{Divis08,Standard12,Liu12}, whereas it confers its electroactivity to a magnon in ErMnO$_3$ \cite{Chaix14}. A puzzling issue remains in the mutual orthogonal orientation of the \m\ and \y\ magnetic moments which excludes a coupling mechanism by isotropic exchange interactions and calls for more subtle mechanisms. The deviation of the $x_{Mn}$ coordinate from 1/3 has also been proposed as an important parameter in the selection of the \m\ magnetic configuration, either by triggering the sign of the inter-layer \m\ magnetic effective interaction \cite{Fabreges09}, or by determining the orientation of the \m\ within the ($a$, $b$) plane through spin-lattice coupling \cite{Solovyev12}. In our study, this parameter does not seem to vary significantly with the temperature, impeding a definite conclusion on this issue.

 
\noindent {\bf Acknowledgments}\\
This work was partly supported by the French ANR Project DYMAGE (ANR-13-BS04-0013) and by the HLD at HZDR, member of the European Magnetic Field Laboratory (EMFL). DZD gratefully acknowledges the financial support by the project DN – 08/9 of the Bulgarian Science Fund. The work of M. I. A. was supported by the Government of the Basque Country (Project No. IT779-13) and the Spanish Ministry of Economy and Competitiveness and FEDER funds (Project No. MAT2015-66441-P). A. A. M. and V. Yu. I. gratefully acknowledge the financial support by the Program of Russian  Academy of Sciences ``Actual problems of low temperature physics''.

\end{document}